\documentclass{iopart}
\usepackage{iopams}
\usepackage{graphicx,epsf}
\usepackage{color}
\usepackage{mathrsfs}
\usepackage{dcolumn}
\usepackage{cite}

\usepackage{subfig}

\begin{document}


\title[]{Zero frequency zonal flow excitation by energetic electron driven beta-induced Alfv\'en eigenmode}

\author{Zhiyong Qiu$^{1}$, Liu Chen$^{1,2}$, Fulvio Zonca$^{3,1}$ and Ruirui Ma$^4$}

\address{$^1$Institute for    Fusion Theory and Simulation and Department of Physics, Zhejiang University, Hangzhou, P.R.C}
\address{$^2$Department of   Physics and Astronomy,  University of California, Irvine CA 92697-4575, U.S.A.}
\address{$^3$ ENEA, Fusion and Nuclear Safety Department, C. R. Frascati, Via E. Fermi 45, 00044 Frascati (Roma), Italy}
\address{$^4$ Southwestern Institute of Physics - P.O. Box 432 Chengdu 610041, P.R.C.}

\begin{abstract}
Zero frequency zonal flow (ZFZF) excitation by trapped energetic electron driven beta-induced Alfv\'en eigenmode (eBAE) is investigated using nonlinear gyrokinetic theory. It is found that, during the linear growth stage of eBAE,  resonant energetic electrons (EEs) not only effectively drive eBAE unstable, but also  contribute to the nonlinear coupling, leading to ZFZF excitation. The trapped EE contribution to ZFZF generation is dominated by EE responses to eBAE in the ideal region, and is comparable to thermal plasma contribution to Reynolds and Maxwell stresses.
\end{abstract}

\maketitle

\section{Introduction}\label{sec:intro}

Energetic particle (EP) related physics are expected to play crucial roles in burning plasmas of future reactors \cite{IPBNF1999,AFasoliNF2007,LChenRMP2016}. Energetic fusion alpha particles   heating of fuel ions through  collisional  as well as collisionless channels \cite{NFischPRL1992,NFischNF1994} is crucial for  achieving   self-sustained burning. On the other hand,  free energy associated with EPs pressure gradient,  may drive collective instabilities, e.g., shear Alfv\'en waves (SAWs)   \cite{AFasoliNF2007,LChenRMP2016}, and induce EP anomalous transport by wave-particle interactions \cite{LChenJGR1999}. Thus, for quantitative understanding of plasma confinement and fusion performance, in-depth understanding of SAW instabilities   including saturation based on  first-principle-based theory is needed \cite{LChenRMP2016,FZoncaNJP2015}.
Among the various nonlinear saturation mechanisms,   excitation of zero frequency zonal fields \cite{MRosenbluthPRL1998,LChenPRL2012,ZQiuNF2017} is an important route \cite{LChenPoP2013}. Zonal fields, including zonal flow (ZF) and zonal current (ZC), are toroidally and  predominantly poloidally symmetric structures \cite{PDiamondPPCF2005}; corresponding to the nonlinear equilibria of plasma in the presence of finite amplitude fluctuations such as drift wave (DW) and/or drift Alfv\'en wave  (DAW)  turbulence \cite{LChenNF2007a,MFalessiPoP2019}. Zonal fields are nonlinearly excited by DW/DAW turbulence through modulational instability, and in turn, scatter DW/DAW turbulence into linearly stable short wavelength   regime \cite{ZLinScience1998,LChenPoP2000,LChenPRL2012}. Up to now, most theoretical investigations on zonal fields generation have been focused on toroidal Alfv\'en eigenmodes (TAEs)  \cite{CZChengAP1985} as  proof of principle  demonstration \cite{LChenPRL2012,ZQiuPoP2016,ZQiuEPL2013,ZQiuNF2017,ZQiuPRL2018,YTodoNF2010}. Extensions  to other SAW instabilities, e.g., beta-induced Alfv\'en eigenmode (BAE) \cite{WHeidbrinkPRL1993,FZoncaPPCF1996} and/or reversed shear Alfv\'en eigemode \cite{FZoncaPoP2002}, are straightforward \cite{ZQiuNF2016,SWei2020}.

Nonlinear excitation of zonal field by TAE was investigated using gyrokinetic theory in Ref. \cite{LChenPRL2012}, and  it was shown that  zonal field generation is the result of breaking of pure Alfv\'enic state by toroidicity \cite{LChenPoP2013}. Here, ZC generation is favored due to its lower threshold, while ZF generation is shielded by  neoclassical polarization effects \cite{MRosenbluthPRL1998,LChenPRL2012}. It was further shown that,  during the linear  growth stage of TAE, the resonant EPs that drive   TAE unstable  also contribute to and dominate the zonal field generation process; which renders the zonal field generation into a forced driven process.  The EP contribution in the ideal region, meanwhile,  overcomes the thermal plasma contribution to Reynolds stress (RS) and Maxwell stress (MS)  in the fast radially varying inertial region \cite{ZQiuPoP2016,YTodoNF2010}. A peculiar feature of zonal field generated by Aflv\'en eigenmodes (AEs)  is that, due to the fine radial structure of AEs, the nonlinearly generated zonal field  also has a micro-scale radial structure  \cite{ZQiuNF2016,ZQiuNF2017};  in addition to the usual well-known meso-scale structure \cite{ZLinScience1998,LChenPoP2000,PDiamondPPCF2005}. This additional fine-scale radial structure  may enhance the nonlinear coupling and the regulation effects of zonal fields on TAE; consequently, it leads to lower TAE saturation level  and is, thus, important in quantitative prediction of EP transport.

Among various AEs, beta-induced Alfv\'en eigenmode (BAE) \cite{WHeidbrinkPRL1993,FZoncaPPCF1996}, excited in the low frequency continuum gap induced by plasma compression and diamagnetic effects \cite{XWangPoP2011}, is of particular interest. BAEs, with their relatively low frequency $(\omega\sim O(v_i/R_0))$, can interact with and be driven unstable by the free energy associated with both thermal ions as well as EPs,  in different wavelength regimes. Here, $v_i$ is the ion thermal velocity and $R_0$ is the major radius.  It is observed in HL-2A tokamak that  the BAEs can also be driven unstable by energetic electrons (EE) generated   in both Ohmic and electron cyclotron resonance heating (ECRH) plasmas \cite{XDingNF2002,WChenPRL2010}; and it is found that  the condition for this EE-driven BAE (eBAE) destabilization  is related to not only the population, but also the pitch angle and energy of EEs. It is shown, in both gyrokinetic simulations using HL-2A parameters \cite{JChengPoP2016,JChengPoP2017},  as well as gyrokinetic analytical theory based on generalized fishbone like dispersion relation \cite{RMaNF2020,FZoncaPoP2014a,FZoncaPoP2014b}, that  eBAE can be driven unstable by the precessional resonance of trapped EEs.
Nonlinear generation of zonal field by BAE, on the other hand, is investigated using both gyrokinetic simulation \cite{HZhangPST2013}  as well as gyrokinetic theory \cite{ZQiuNF2016}. Due to the flute like mode structure with $|k_{\parallel}qR_0|\sim \sqrt{\beta}$, zonal field generation is dominated by electro-static ZF due to thermal ion RS \cite{ZQiuNF2016}.  Here, $k_{\parallel}$ is the parallel wavenumber, $q$ is the safety factor and $\beta$ is the plasma thermal to magnetic pressure ratio. It is, thus, natural to expect that  resonant EEs  could also contribute to zero frequency zonal flow (ZFZF) generation in the linear growth stage of eBAE \cite{ZQiuPoP2016}, and this constitutes the main motivation of the present work.

In this work, nonlinear ZFZF generation during the linear growth stage of eBAE is investigated, with emphasis on the contribution of resonant EEs to the nonlinear coupling. The rest of the paper is organized as follows. In Sec. \ref{sec:model}, the theoretical model is given. The linear particle responses to eBAE is derived in Sec. \ref{sec:linear_eBAE_dr}, which is then applied to derive the linear WKB dispersion relation of eBAE. The nonlinear excitation of ZFZF by eBAE, with the contribution of both thermal plasmas and resonant EEs, is investigated in Sec. \ref{sec:zfzf_generation}. And finally, a brief summary is given in Sec. \ref{sec:conclusion}.

\section{Theoretical model}\label{sec:model}

For the clarity of analysis while focusing on the physics picture of  EE effect on ZFZF generation by eBAE, we consider a simple tokamak equilibrium with circular magnetic surfaces, and the equilibrium magnetic field is given as $\mathbf{B}=B_0(\mathbf{e}_{\xi}/(1+\epsilon\cos\theta)+(\epsilon/q)\mathbf{e}_{\theta})$. Here, $\epsilon\equiv r/R_0\ll1$ is the inverse aspect ratio of the torus, $r$ is the minor radius, and $\xi$ and $\theta$ are respectively, the toroidal and poloidal angles, forming a toroidal flux coordinate system $(r,\theta,\xi)$. Another assumption to simplify the analysis  is that eBAE is driven unstable by the precessional frequency resonance of deeply trapped EEs \cite{RMaNF2020}. This assumption is reasonable as the typical EEs  transit/bounce frequencies  are too high to resonate with BAE and  the population of barely trapped/passing EEs with lower bounce/transit frequencies  is too small to drive  eBAE unstable.

To investigate ZFZF generation by eBAE with predominantly SAW polarization, $\delta\phi$ and $\delta\psi\equiv \omega\delta A_{\parallel}/(ck_{\parallel})$ are adopted as field variables. Here, $\delta\phi$ and $\delta A_{\parallel}$ are, respectively, the scalar potential and the parallel (to equilibrium magnetic field) vector potential,   and ideal MHD constraint is recovered by taking $\delta\psi=\delta\phi$.  For the nonlinear interactions between ZFZF and eBAE, we have $\delta\phi=\delta\phi_Z+\delta\phi_B$, with $\delta\phi_B=\delta\phi_0+\delta\phi_{0^*}$. Here, the subscripts $Z$, $B$ denote ZFZF and  eBAE,  $\delta\phi_0$ is the pump eBAE and $\delta\phi_{0^*}$ is its complex conjugate.
The well-known ballooning-mode decomposition is assumed:
\begin{eqnarray}
\delta\phi_0=A_0 e^{i(n\xi-m_0\theta-\omega_0t)}\sum_je^{-ij\theta}\Phi_0(x-j).
\end{eqnarray}
Here, $n$ is the toroidal   mode number, $m=m_0+j$ is the poloidal mode number with $m_0$ being its reference value satisfying $nq(r_0)=m_0$, $r_0$ is the radial coordinate where the eBAE is assumed to be localized, $x=nq-m_0\simeq nq'(r_0)(r-r_0)$, $\Phi_0$ is the micro-scale structure due to $k_{\parallel}$ radial dependence as well as magnetic shear, and $A_0\equiv \hat{A}_0 e^{i\int \hat{k}_r dr}$ is the radial envelope with $\hat{k}_r$ being the typically meso-scale radial envelope wavenumber and $\hat{A}_0$ is the envelope amplitude. Thus, we have  $k_r=\hat{k}_r-i\partial_r\ln\Phi_0$ with $|\partial_r\ln\Phi_0|$ typically much larger than $\hat{k}_r$ \cite{ZQiuNF2016}.
On the other hand, for ZFZF, we have
\begin{eqnarray}
\delta\phi_Z=\hat{A}_Ze^{i(\int \hat{k}_Z dr -\omega_Z t)}\sum_j\Phi_Z(x-j),
\end{eqnarray}
with $\Phi_Z$ accounting for the fine radial structure of ZFZF \cite{ZQiuNF2017}, as a result of the micro-scale structure of the pump eBAE. The summation over $j$ in the expression of $\delta\phi_Z$ indicates that the fine structures of $\delta\phi_Z$ locate  at the radial position of $\Phi_0(x-j)$.

One governing nonlinear equation  describing the ZFZF excitation by eBAE, can be  derived from quasi-neutrality condition
\begin{eqnarray}
\frac{n_0e^2}{T_i}\left(1+\frac{T_i}{T_e}\right)\delta\phi_k=\sum_s \left\langle q_s J_k\delta H_k \right\rangle_s.\label{eq:QN}
\end{eqnarray}
Here, $\langle\cdots\rangle$ indicates velocity space integration, $\sum_s$ is the summation over different particle species with $s=i,e,h$ denoting ions, electrons and EEs, respectively, $J_k\equiv J_0(k_{\perp}\rho)$ with $J_0$ being the Bessel function of zero index accounting for finite Larmor radius (FLR) effects, $\rho=v_{\perp}/\Omega_c$ is the Larmor radius, $\Omega_c=q_sB/(mc)$ is the cyclotron frequency and, meanwhile, $\delta H_k$ is the non-adiabatic part of particle response,   which can be derived from the nonlinear gyrokinetic equation \cite{EFriemanPoF1982}:
\begin{eqnarray}
&&\left(-i\omega+v_{\parallel}\partial_l+i\omega_d\right)\delta H_k=-i\frac{q_s}{m}QF_0J_k\delta L_k \nonumber\\
&&\hspace*{4em}-\Lambda^{k}_{k'',k'}J_{k'}\delta L_{k'}\delta H_{k''}\label{eq:NLGKE}.
\end{eqnarray}
Here, $l$ is the  length along the equilibrium magnetic field line, $QF_0=(\omega\partial_E-m\omega_*/T)F_0$ with $E=v^2/2$ and $F_0$ being the equilibrium particle distribution function,   $\omega_*F_0=T\mathbf{k}\cdot\mathbf{b}\times\nabla F_0/(m\Omega)$ is the diamagnetic frequency,    $\omega_d=(v^2_{\perp}+2 v^2_{\parallel})/(2 \Omega R_0)\left(k_r\sin\theta+k_{\theta}\cos\theta\right)$ is the magnetic drift frequency,  $\delta L_k\equiv\delta\phi_k-k_{\parallel} v_{\parallel}\delta\psi_k/\omega_k$, $\Lambda^{k}_{k'',k'}\equiv (c/B_0)\sum_{\mathbf{k}=\mathbf{k}'+\mathbf{k}''} \mathbf{b}\cdot\mathbf{k''}\times\mathbf{k'}$; and other notations are standard.   The other field equation, the nonlinear gyrokinetic vorticity equation  \cite{LChenJGR1991,FZoncaPoP2014a,LChenRMP2016}  is derived from parallel Amp\'ere's law and nonlinear gyrokinetic equation as
\begin{eqnarray}
&&\frac{c^2}{4\pi \omega^2_k}B\frac{\partial}{\partial l}\frac{k^2_{\perp}}{B}\frac{\partial}{\partial l}\delta \psi_k +\frac{e^2}{T_i}\left\langle (1-J^2_ k)F_0\right\rangle\delta\phi_k\nonumber\\
&&-\sum_s\left\langle\frac{q}{\omega_k}J_k\omega_d\delta H_k \right\rangle_s\nonumber\\
&=&-i \sum_s\left\langle q\Lambda^{k}_{k'',k'}J_k\delta L_{k'}\delta H_{k''} \right\rangle_s.\label{eq:vorticityequation1}
\end{eqnarray}
The terms on the left hand side  (LHS) of equation (\ref{eq:vorticityequation1}) are, respectively, field line bending term, inertial term and curvature coupling term. The term on the right hand side (RHS) is the formally  nonlinear term, which can be further simplified using quasi-neutrality condition, equation (\ref{eq:QN}).
Adding $i\Lambda^{k}_{k'',k'} \delta\phi_{k'} \left[(n_0e^2/T_i)(1+T_i/T_e) \delta\phi_{k''}-\sum_s\langle qJ_{k''}\delta H_{k'',s}\rangle \right]$  on the right hand side of equation (\ref{eq:vorticityequation1}), the vorticity equation reduces to  the expression we are familiar with \cite{FZoncaPoP2004,LChenRMP2016}:
\begin{eqnarray}
&&\frac{c^2}{4\pi \omega^2_k}B\frac{\partial}{\partial l}\frac{k^2_{\perp}}{B}\frac{\partial}{\partial l}\delta \psi_k +\frac{e^2}{T_i}\left\langle (1-J^2_ k)F_0\right\rangle\delta\phi_k\nonumber\\
&&-\sum_s\left\langle\frac{q}{\omega_k}J_k\omega_d\delta H_k \right\rangle_s\nonumber\\
&=&-i\frac{\Lambda^{k}_{k'',k'}}{\omega_k}\left [ \frac{c^2}{4\pi}k''^2_{\perp} \frac{\partial_l\delta\psi_{k'}\partial_l\delta\psi_{k''}}{\omega_{k'}\omega_{k''}} \right.\nonumber\\
&&\left.+ \sum_s\left\langle e(J_kJ_{k'}-J_{k''})\delta L_{k'}\delta H_{k''}\right\rangle \right].
\label{eq:vorticityequation}
\end{eqnarray}
The terms  on the right hand side (RHS) of equation (\ref{eq:vorticityequation}) are the  Maxwell and Reynolds stresses (MX\&RS) \footnote{Interested readers may refer to Ref. \citenum{ZQiuNF2019a} (footnote 8 therein) for a brief discussion of the derivation of equation (\ref{eq:vorticityequation}).}.  Note that, in addition to the usual RS\&MX, the EE nonlinearity  will also contribute to  the curvature coupling term through the nonlinear EE response to ZFZF \cite{ZQiuPoP2016,RMaNF2020}. The reason that both equations  (\ref{eq:vorticityequation1}) and (\ref{eq:vorticityequation}), which are equivalent to each other, are introduced   is that  equation (\ref{eq:vorticityequation}) is used in Ref. \cite{ZQiuNF2016} to investigate the spontaneous ZFZF generation by BAE via modulational instability, where the   contribution of thermal plasma to the nonlinear coupling is derived conveniently from RS\&MX. That result will be readily generalized to the self-coupling of eBAEs in the present work. Meanwhile, as we will show later, it is easier to derive the EE contribution  using equation (\ref{eq:vorticityequation1}).
Since the resonant EEs effect of interest here   can only be important during the linear growth stage of eBAE,  the ZFZF generation is a forced driven process \cite{ZQiuPoP2016,ZQiuNF2017,ABiancalaniPoP2020}, and the feedback of ZFZF to eBAE will not be considered here.

\section{Linear WKB dispersion relation of eBAE}\label{sec:linear_eBAE_dr}

The linear particle responses to eBAE  will be derived  and applied for deriving the nonlinear   responses to ZFZF. Here, the sketched derivation of eBAE WKB dispersion relation will also be presented, while the more quantitative analysis of its stability, is much more straightforwardly carried out in ballooning space, under the framework of general fishbone like dispersion relation \cite{RMaNF2020,FZoncaPoP2014a,FZoncaPoP2014b}.

The thermal electron response to eBAE, noting $\omega_B\sim v_i/R_0\ll k_{\parallel}v_e$, is given as
\begin{eqnarray}
\delta H_{e,0}=-\frac{e}{T_e}F_0\delta\psi_0.\nonumber
\end{eqnarray}
The thermal ion response to eBAE, can be derived order by order, using $k_{\perp}\rho_i$ and $1/q$ as the smallness parameters. To the lowest order, one has
\begin{eqnarray}
\delta H^{(0)}_{i,0}=\frac{e}{T_i}F_0J_0\delta\phi^{(0)}_0.\nonumber
\end{eqnarray}
Substituting into quasi-neutrality condition, one has,   $\delta\phi^{(0)}_0=\delta\psi^{(0)}_0$, i.e., ideal MHD is maintained in the leading order. To the next order, one has,
\begin{eqnarray}
\delta H^{(1)}_{i,0}=\frac{e}{T_i}F_0J_0\left(\delta\phi^{(1)}_0+\frac{\omega_{d,i}}{\omega_0}\delta\phi^{(0)}_0\right),\nonumber
\end{eqnarray}
with
\begin{eqnarray}
\delta\phi^{(1)}_0=\tau\left\langle\frac{F_0}{n_0}\frac{\omega_{d,i}}{\omega_0}\right\rangle\delta\phi^{(0)}_0\nonumber
\end{eqnarray}
derived from quasi-neutrality condition and $\tau\equiv T_e/T_i$. The WKB dispersion relation of SAW, with the effect of thermal ion compression induced continuum upshift \cite{XWangPoP2011,FZoncaPPCF1996}, can be derived by substituting the thermal plasma responses into the vorticity equation.

The nonadiabatic trapped EE response to eBAE can be derived, noting the $k_{\parallel}v_e\sim \sqrt{\epsilon}v_{h}/R_0\gg\omega,\omega_{d,h}$ ordering, and to the leading order, we have
\begin{eqnarray}
\delta H_h=\frac{e}{m}\frac{QF_{0,h}}{\omega}\delta\psi+\delta K_h,
\end{eqnarray}
with $\delta K_h$ derived from
\begin{eqnarray}
\left(-i\omega+v_{\parallel}\partial_l+i\omega_d\right)\delta K_h=-i\frac{e}{m}\frac{QF_{0,h}}{\omega}\left(\delta\phi-\delta\psi+\frac{\omega_{d,h}}{\omega}\delta\phi\right).\label{eq:QK_EE_linear}
\end{eqnarray}
The third term   on the RHS of equation (\ref{eq:QK_EE_linear})  dominates, as $\delta\phi-\delta\psi\sim (\omega_{d,i}/\omega)\delta\phi\ll (\omega_{d,h}/\omega)\delta\phi$ and that magnetic drift frequency $\omega_d$ scales with particle characteristic energy.  Noting again that $k_{\parallel}v_{h}\gg\omega_B, \omega_{d,h}$, one then has, $\delta K_h$ is independent of $l$ to the leading order. Thus,   $\delta K_h=\delta G_h \exp(in(\xi-q\theta))$ with $\delta G_h$ being independent of $l$, and equation (\ref{eq:QK_EE_linear}) reduces to
\begin{eqnarray}
(-i\omega+i\omega_d)\delta G_h=-i\frac{e}{m}\frac{QF_{0,h}}{\omega}\omega_d\delta\phi_0^{(0)} e^{-in(\xi-q\theta)}.
\end{eqnarray}
Taking $\delta\phi_0^{(0)}\simeq e^{i(n \xi-m\theta)}\hat{\psi}$, we then have
\begin{eqnarray}
\delta G_h=\frac{e}{m}\frac{QF_{0,h}}{\omega}\frac{\Omega_d}{\omega-\overline{\overline{\omega}}_d}\hat{\psi},
\end{eqnarray}
with $\Omega_d\equiv \overline{\overline{\omega_d e^{i(nq-m)\theta}}}\simeq \overline{\overline{\omega}}_d$ for BAEs   typically localized around rational surfaces, $\overline{\overline{(\cdots)}}\equiv \tau^{-1}_b \oint (dl/v_{\parallel})(\cdots)$ denoting bounce average, and  $\tau_b\equiv \oint dl/v_{\parallel}$ is the magnetically  trapped particle bounce period.
The trapped EE response to BAE  can be written as
\begin{eqnarray}
\delta H_h=\frac{e}{m}\frac{QF_{0,h}}{\omega}\delta\psi_0^{(0)}+ \frac{e}{m}\frac{QF_{0,h}}{\omega}\frac{\Omega_d}{\omega-\overline{\overline{\omega}}_d}\hat{\psi}e^{in(\xi-q\theta)}.
\end{eqnarray}
Substituting $\delta H_h$ into the vorticity equation, noting that the flux surface average of a velocity space integral can be written as  $\overline{\langle\cdots\rangle} = \sum_{\sigma}  \int E dEd\Lambda (\tau_b/(2\pi qR_0))\overline{\overline{(\cdots)}}$,  with $\sigma$ the sign of $v_{\parallel}$ and $\Lambda=\mu B_0/E$ the dimensionless pitch angle variable,  we then have  the WKB dispersion relation of eBAE
\begin{eqnarray}
\frac{n_0e^2}{T_i}\hat{b}_i \mathscr{E}_{eBAE}\delta\phi_0^{(0)}=0,
\end{eqnarray}
with $\mathscr{E}_{eBAE}$ defined as
\begin{eqnarray}
\mathscr{E}_{eBAE}\equiv &-&\frac{k^2_{\parallel}V^2_A}{\omega^2}+1-\frac{\omega^2_G}{\omega^2}\nonumber\\
&+&\frac{T_i}{2\pi n_0m_eqR_0\hat{b}_i\omega^2}\sum_{\sigma=\pm1}\int EdEd\Lambda \tau_b QF_{0,h}\frac{\Omega^2_d}{\omega-\overline{\overline{\omega}}_d}.\label{eq:eBAE_DR}
\end{eqnarray}
Here,  $\hat{b}_i  =  k^2_{\perp}\rho^2_i/2$  with $\rho^2_i=2 T_i/(m_i\Omega^2_i)$.  The eBAE eigenmode dispersion relation can be derived by asymptotic matching of the radially fast varying inertial region with the slowly varying ideal region \cite{FZoncaPoP2014a}. Interested readers may refer to Ref. \cite{RMaNF2020} for the detailed discussion of the linear eBAE stability, including the dependence on EE distribution function. Note that, in  equation (\ref{eq:eBAE_DR}), the last term accounting for resonant EE drive has a ``$\hat{b}_i$" in the denominator, which is dominated by $k^2_r\rho^2_i/2$ in the radially fast varying inertial layer; while the numerator  has no dependence on the fast varying mode structures. Thus, the trapped EE contribution is mainly in  the ideal region \cite{RMaNF2020}, even though the EE drift/banana orbit width is typically smaller than inertial layer width. This is due to the fact that, EE bounce averaged response is governed by normal curvature, as noted in Refs. \cite{IChavdarovskiPPCF2009,IChavdarovskiPoP2014},  while thermal ion response to the underlying BAE-like fluctuation is due to geodesic curvature  \cite{FZoncaPPCF1996}.

\section{Nonlinear ZFZF generation by eBAE}\label{sec:zfzf_generation}

It has been shown that, electrostatic ZF generation dominates due to the $\omega\ll\omega_A$ frequency range of BAE \cite{ZQiuNF2016}, and EP effect is, typically, important in its contribution to energy density in the vorticity equation rather than the particle density in the  quasi-neutrality condition \cite{LChenPS1995,ZQiuPoP2016}.
Thus, ZFZF generation by eBAE  can be derived by substituting the particle responses into the surface averaged nonlinear vorticity equation,
\begin{eqnarray}
&&\frac{e^2}{T_i}\langle(1-J^2_Z)F_0\rangle\overline{\delta\phi_Z}-\sum_{s=e,i}\left\langle{\overline{\frac{q}{\omega}J_Z\omega_d\delta H^L_Z}}\right\rangle\nonumber\\
&=& -\frac{i}{\omega}\Lambda^{k}_{k'',k'}\left [ \frac{c^2}{4\pi}k''^2_{\perp} \frac{\partial_l\delta\psi_{k'}\partial_l\delta\psi_{k''}}{\omega_{k'}\omega_{k''}} \right.\nonumber\\
&&\left.+ \sum_s\left\langle e(J_ZJ_{k'}-J_{k''})\delta L_{k'}\delta H_{k''}\right\rangle \right] +\left\langle\overline{\frac{q}{\omega}J_Z\omega_d\delta H^{NL}_{Z,h}}\right\rangle.\label{eq:ZF_vorticity}
\end{eqnarray}
Here, the linear and nonlinear EE responses to ZFZF are formally separated; i.e., $\delta H_{Z,h}=\delta H^L_{Z,h}+\delta H^{NL}_{Z,h}$, and the nonlinear terms are formally written on the right hand side.

Thermal plasma contribution to RS\&MX, can be derived following Ref. \cite{ZQiuNF2017,ZQiuNF2016}\footnote{Interested readers may refer to equation (8) of Ref. \cite{ZQiuNF2017} and equation (4) of Ref. \cite{ZQiuNF2016} for the derivation.}, and we have
\begin{eqnarray}
\hspace*{-4em}\mbox{RS+MX}=&-&\frac{1}{2}\frac{c}{B_0}\frac{n_0e^2}{T_i}k_{\theta}\rho^2_i\frac{1}{\omega_Z}\left(1-\frac{k^2_{\parallel,0}V^2_A}{\omega^2}\right)\frac{\partial^2}{\partial r^2}\left(\hat{F}|\hat{A}_0|^2\sum_m |\Phi_0|^2\right).\label{eq:RSMX}
\end{eqnarray}
Here, $\hat{F}=i(\hat{k}_{r,0}-\hat{k}_{r,0^*})+\partial_r\ln\Phi_0-\partial_r\ln\Phi_{0^*}$ with $\hat{k}_{r,0}-\hat{k}_{r,0^*}$ accounting for radial envelope modulation  and $\partial_r\ln\Phi_0-\partial_r\ln\Phi_{0^*}$ related with fine radial structures of BAE \cite{ZQiuNF2016}.  For AEs characterized with fine radial structure due to SAW continuum coupling, we typically have $\hat{F}\simeq  \partial_r\ln\Phi_0-\partial_r\ln\Phi_{0^*}$.  The occurrence of $\partial^2_r$ in the above expression  indicates that thermal plasma nonlinearity dominates as BAE mode structure is radially fast varying, i.e., in the inertial layer of AEs. On the other hand, for BAEs with mode structures typically localized near the accumulation point of the BAE gap, the net contribution of MX (the  term proportional to $k^2_{\parallel}V^2_A/\omega^2$ in equation (\ref{eq:RSMX})) is  much smaller than the other term, and $\langle\langle1-k^2_{\parallel}V^2_A/\omega^2\rangle\rangle_0\simeq 1$ with $\langle\langle\cdots\rangle\rangle_0$ being the average over eBAE mode structure.

The effects of resonant EE to ZFZF generation by eBAE are the focus of the present work. Nonlinear effects of EE  can enter through both the curvature coupling term via the nonlinear   EE response to ZFZF, as well as the  EE contribution to RS and MX terms \footnote{Here, for EE contribution to RS and MX, we mean the nonlinearity related to $\delta \mathbf{v}\cdot\nabla\delta\mathbf{v}$ and $\delta\mathbf{j}\times\delta\mathbf{B}$ terms in MHD momentum equation. For eBAE discussed here, these two   terms are dominated by their ideal region contribution, but we still call them  ``EE contribution to RS\&MX" for convenience. }. This is one major difference with energetic ion contribution to ZFZF generation by TAE investigated in Ref. \cite{ZQiuPoP2016}, where only resonant energetic ion contribution to curvature coupling term need to be considered due to the large energetic ion drift orbit size.
In fact, as we will demonstrate, EEs contributions to the  RS\&MX  (again, here we mean the nonlinear terms related  to $\delta \mathbf{v}\cdot\nabla\delta\mathbf{v}$ and $\delta\mathbf{j}\times\delta\mathbf{B}$) terms in the ideal region dominate over the EE curvature coupling terms; even though the EE orbit size is typically smaller or comparable to the inertial layer width of BAEs $\Delta r\sim \sqrt{\beta}/(nq')$.

We start from the resonant EE contribution to curvature coupling term.   Nonlinear trapped EE response to ZFZF, can be derived from nonlinear gyrokinetic quation
\begin{eqnarray}
\left(-i\omega+\omega_{tr,h}\partial_{\theta}+i\omega_d\right)\delta H^{NL}_{Z,h}=-\Lambda^{k}_{k'',k'}\delta L_{k'}\delta H_{k''}.
\end{eqnarray}
Taking $\delta H^{NL}_{Z,h}=e^{i\Lambda_Z}\delta H^{NL}_{dZ,h}$  with $\Lambda_Z$ defined by $\omega_{tr,h}\partial_{\theta}\Lambda_Z+\omega_d-\overline{\overline{\omega}}_d=0$, $e^{i\Lambda_Z}$ is the operator for    banana-center coordinate transform,  and separating d.c. from a.c. responses by letting $\delta H^{NL}_{dZ,h}=\overline{\overline{\delta H^{NL}_{dZ,h}}}+\widetilde{\delta H^{NL}_{dZ,h}}$, we then have
\begin{eqnarray}
-i\omega_Z\overline{\overline{\delta H^{NL}_{dZ,h}}}&=&-\Lambda^{k}_{k'',k'}\overline{\overline{\delta L_{k'}\delta H_{k''} e^{-i\Lambda_Z}}},\label{eq:NL_EE_dc}\\
\hspace*{-4em}\left(-i\omega_Z+\omega_{tr,h}\partial_{\theta}\right)\widetilde{\delta H^{NL}_{dZ,h}}&=&-\Lambda^{k}_{k'',k'}\left[\delta L_{k'}\delta H_{k''} e^{-i\Lambda_Z}-\overline{\overline{\delta L_{k'}\delta H_{k''} e^{-i\Lambda_Z}}}\right].\label{eq:NL_EE_ac}
\end{eqnarray}
Thus, $|\overline{\overline{\delta H^{NL}_{dZ,h}}}/\widetilde{\delta H^{NL}_{dZ,h}}|\sim |\omega_{tr,h}/\omega_Z|\gg1$ and consequently, $\delta H^{NL}_{Z,h}\simeq\overline{\overline{\delta H^{NL}_{dZ,h}}} e^{i\Lambda_Z}$.

Substituting $\delta H^{NL}_{Z,h}$ into the curvature coupling term, noting that $\omega_d=-\omega_{tr}\partial_{\theta}\Lambda_{Z,h}$, and integrating by parts, we then have
\begin{eqnarray}
\left\langle\overline{\frac{q}{\omega}\omega_d\delta H^{NL}_{Z,h}}\right\rangle = i\frac{e}{\omega}\left\langle \overline{e^{i\Lambda_{Z,h}} \omega_{tr,h}\partial_{\theta}\widetilde{\delta H^{NL}_{dZ,h}}}\right\rangle.\label{eq:curvature_coupling}
\end{eqnarray}
Substituting equation (\ref{eq:NL_EE_ac}) into (\ref{eq:curvature_coupling}),  neglecting the odd terms of $v_{\parallel}$ which vanishes in velocity space integration, and keeping only the dominant contribution of resonant EEs from $\delta K_h$, we have
\begin{eqnarray}
&&\left\langle\overline{\frac{q}{\omega}\omega_d\delta H^{NL}_{Z,h}}\right\rangle\nonumber\\
&=& -i\frac{e}{\omega}\left\langle\overline{\left(e^{i\Lambda_{Z,h}}-\overline{\overline{e^{i\Lambda_{Z,h}}}}\right) \Lambda^{k}_{k'',k'} \delta\phi_{k'}\delta K_{k''} e^{-i\Lambda_{Z,h}}}\right\rangle\nonumber\\
&\simeq& -\frac{e}{\omega}\frac{c}{B_0}\left\langle \overline{\left(1-J^2_0(\hat{\Lambda}_{Z,h})\right) k_{\theta}\frac{\partial}{\partial r}\left(\delta L_0\delta K_{0^*}-\delta L_{0^*}\delta K_0\right)} \right\rangle.
\end{eqnarray}

For ZFZF generation due to eBAE self-coupling, noting  $Q_{0^*}=-Q_0$, $\omega_{0^*}=-\omega^*_0$, $\Omega_{d0^*}=-\Omega_{d0}$,  and, thus,
\begin{eqnarray}
\frac{1}{\omega_0-\overline{\overline{\omega}}_{d0}}-\frac{1}{\omega^*_{0}+\overline{\overline{\omega}}_{d0^*}}\simeq -2i\pi\delta(\omega_0-\overline{\overline{\omega}}_{d0}). \nonumber
\end{eqnarray}
Furthermore, we note $\delta L_0=\delta\phi_0-k_{\parallel}v_{\parallel}/\omega_0\delta\psi_0=(\overline{\overline{\omega}}_{d,0}/\omega_0)\delta\phi_0$ for resonant EEs; hence,
\begin{eqnarray}
\hspace*{-7em}&&\left\langle\overline{\frac{q}{\omega}\omega_d\delta H^{NL}_{Z,h}}\right\rangle 
\simeq  -2i\pi\frac{c}{B_0}\frac{e^2}{m_e}k_{\theta}\partial_r| \hat{\psi}_0|^2 \left\langle \overline{ (1-J^2_0(\hat{\Lambda}_{Z,h}))Q_0F_{0,h}\frac{\Omega^2_d}{\omega^2_0}\delta(\omega_0-\overline{\overline{\omega}}_{d0})} \right\rangle.\label{eq:CC_final}
\end{eqnarray}
In deriving equation (\ref{eq:CC_final}), we have neglected higher order ($\sim O(|\hat{\Lambda}^2_{Z,h}|)$)  terms.     Note that, the EE orbit is typically smaller than eBAE mode width; i.e.,  $|\hat{\Lambda}_{Z,h}|\ll1$ and $|1-J^2_0(\hat{\Lambda}_{Z,h})|\ll1$, so the contribution in equation (\ref{eq:CC_final}) becomes negligible.  In this parameter regime,  the EE contribution to RS and MX, thus, should also be properly evaluated. The corresponding nonlinear EE contribution, however, is easier to evaluate using the expression of equation (\ref{eq:vorticityequation1}), and we have
\begin{eqnarray}
&&-i\left\langle\overline{ q\Lambda^{k}_{k'',k'}J_k\delta L_{k'}\delta H_{k''} }\right\rangle_h\nonumber\\
&=&i e\frac{c}{B_0}\left\langle \overline{\overline{e^{i\Lambda_{Z,h}}}}\  \times\ \overline{\overline{\overline{e^{-i\Lambda_{Z,h}}}} k_{\theta}(-i\partial_r)\left(\delta L_0\delta K_{0^*}-\delta L_{0^*}\delta K_0\right)} \right\rangle\nonumber\\
&\simeq& 2i\pi \frac{c}{B_0}\frac{e^2}{m_e}k_{\theta}\partial_r|\hat{\psi}_0|^2\left\langle\overline{ J^2_0(\hat{\Lambda}_{Z,h})Q_0F_{0,h} \frac{\Omega^2_d}{\omega^2_0}\delta(\omega_0-\overline{\overline{\omega}}_d)}\right\rangle.\label{eq:EE_RSMX}
\end{eqnarray}
Thus,  in the parameter region with EE drift obit size much smaller than the inertial layer width $\Delta r\sim \sqrt{\beta}/(nq')$,  the EE contribution to RS\&MX dominates over that due to nonlinear curvature coupling. In deriving equation (\ref{eq:EE_RSMX}), $|\hat{\Lambda}_{Z,h}|\lesssim1$ is assumed in taking the surface average.   We remark that,  even though the present analysis cannot be applied directly to energetic ions with the drift orbit size   much bigger than the inertial layer width, one can qualitatively speculate from equations (\ref{eq:curvature_coupling}) and (\ref{eq:EE_RSMX}), that     the curvature coupling term in the ideal region dominates, as discussed in Ref. \cite{ZQiuPoP2016}. Substituting equations  (\ref{eq:RSMX}), (\ref{eq:CC_final}) and (\ref{eq:EE_RSMX}) into equation (\ref{eq:ZF_vorticity}), we then have
\begin{eqnarray}
\hspace*{-2em}\omega_Z \chi_Z\overline{\delta\phi}_Z&=&  \frac{c}{B_0}k_{\theta,0} \left[-\frac{1}{2}\rho^2_i \left(1-\frac{k^2_{\parallel,0}V^2_A}{\omega^2_0}\right) \left(\partial_r\ln\Phi_0-\partial_r\ln\Phi_{0^*}\right)\partial^2_r|\delta\phi_0|^2\right.\nonumber\\
&&\left.+2 i \pi \frac{T_i}{n_0m_e}  \partial_r |\delta \phi_0|^2 \left\langle\overline{ Q_0F_{0,h}\frac{\Omega^2_d}{\omega^2_0}\delta(\omega-\overline{\overline{\omega}}_{d0})}\right\rangle \right]
\end{eqnarray}
with $\chi_Z$ being the well-known neoclassical polarization of ZFZF \cite{MRosenbluthPRL1998}, defined as
\begin{eqnarray}
\chi_Z \overline{\delta\phi_Z}\equiv \left(1-\left\langle \frac{F_0}{n_0} J^2_Z|\overline{\overline{e^{i\Lambda_{Z,i}}}}|^2\right\rangle\right)\overline{\delta\phi_Z},
\end{eqnarray}
and  $\chi_Z\simeq 1.6 k^2_Z\rho^2_i/\sqrt{\epsilon}$ in the long wavelength limit \cite{MRosenbluthPRL1998}.

Taking advantage of the radial scale separation,  it is natural to take $\Phi_Z=|\Phi_0|^2$ and $|\hat{k}_Z|\ll |\partial_r\ln\Phi_Z|\sim\sqrt{\beta}/(nq')$. Noting  that $|\omega_{*,h}|>\omega_0$ for eBAE destabilization and, hence, $QF_{0,h}\simeq -\omega_{*,h}F_{0,h}$,  we have
\begin{eqnarray}
\hspace*{-5em} \omega_Z\chi_ZA_Z=\frac{i}{2}\frac{c}{B_0}k_Zk_{\theta,0}\left[k^2_Z\rho^2_i -4i\pi \frac{n_h}{n_0} \left\langle \overline{\frac{\omega_{*,h} F_{0,h}}{n_h} \frac{\Omega^2_d}{\omega^2_0}\delta(\omega-\overline{\overline{\omega}}_{d0})} \right\rangle    \right]|\hat{A}_0|^2. \label{eq:ZF_final}
\end{eqnarray}
It is clear that  the first term on the right hand side of equation (\ref{eq:ZF_final})  has the typically expression  (i.e., $\propto k_{\theta}k^3_Z$ due to polarization nonlinearity)  of ZFZF generation by DWs turbulence \cite{LChenPoP2000} and/or DAW \cite{LChenPRL2012}, with the usual meso-scale radial envelope $\hat{k}_Z$ replaced by the fine-scale structure wave nunmber $k_Z$. Meanwhile, the second term, with an expression closely related to the EE resonant drive of eBAE  as clearly seen by comparing with   the eBAE WKB dispersion relation (\ref{eq:eBAE_DR}), has a much weaker ($\propto k_Zk_{\theta}$) dependence on the radial variation.  Its dominant contribution is from  the ideal region, as we have  discussed below equation (\ref{eq:eBAE_DR}).
One can estimate that  the two terms on the right hand side of equation (\ref{eq:ZF_final}) are comparable to each other, and, thus, the  resonant EEs and thermal plasma contributes together to ZFZF generation. In comparing the two terms, $k^2_Z\sim\beta^{-1}k^2_{\theta,0}$ is assumed \cite{RMaNF2020}, and the linear dispersion relation of eBAE is used.    Note that, in Ref. \cite{ZQiuPoP2016}, where resonant ion contribution to ZFZF generation by TAE is estimated,  the resonant ions effects in the ideal region  dominate  over the thermal plasma contribution to RS\&MX. The difference  is related to the non-cancellation of thermal plasma RS\&MX, due to the $k_{\parallel}V_A\ll\omega$ orderings in the BAE/eBAE case, as well as the different parameter regimes of EP orbit compared to ZFZF radial scale.

In the linear growth stage of eBAE, $\omega_Z=2i\gamma_L$ with $\gamma_L$ being the linear growth rate of eBAE, and one has
\begin{eqnarray}
\hspace*{-2em}A_Z=\frac{k_Zk_{\theta,0}}{4\gamma_L\chi_Z}\frac{c}{B_0}\left[k^2_Z\rho^2_i -4i\pi \frac{n_h}{n_0} \left\langle \overline{\frac{\omega_{*,h} F_{0,h}}{n_h} \frac{\Omega^2_d}{\omega^2_0}\delta(\omega-\overline{\overline{\omega}}_{d0})} \right\rangle    \right]|\hat{A}_0|^2.
\end{eqnarray}
This is the expression of the ZFZF amplitude driven by eBAE in the linear growth stage. In quantitative estimating the ZFZF amplitude and comparing with numerical results, e.g., Ref. \cite{ABiancalaniPoP2020},   $k_Z\equiv -2i\partial_r\ln\Phi_0\sim O(\sqrt{\beta} k_{\theta,0})$ should be used.

\section{Conclusions and Discussions}\label{sec:conclusion}

In conclusion,   it is found that  the resonant EEs  also contribute significantly to the nonlinear excitation of ZFZF by eBAE  through both the RS\&MX  as well as the curvature coupling terms. Since resonant EE effects can only be important in the linear growth stage of eBAE, the ZFZF generation is a forced driven process, with the ZFZF growth rate being twice of eBAE  growth rate. The dependence of ZFZF amplitude on eBAE amplitude is also derived. 
 
One interesting   point is that, due to the relatively small orbit size of EEs (comparing to that of often studied energetic ions), the ratio of RS\&MX to curvature coupling term is $(J^2_0(\hat{\Lambda}_{Z,h})1/|1-J^2_0(\hat{\Lambda}_{Z,h})|$, with $|\hat{\Lambda}_{Z,h}|$ being the EE drift/banana orbit width comparing to ZFZF micro- radial scale, associated with the eBAE inertial layer width $\Delta r\sim \sqrt{\beta}/(nq')$. For $|\hat{\Lambda}_{Z,h}|\ll1$, the EE contribution to RS\&MX dominates over the curvature coupling term; in contrary to the usually studied energetic ions with the orbit size much larger than inertial layer width and the effects to RS\&MX are typically negligible. The overall EE effect, on the other hand, is comparable to that of thermal plasma, and is less significant than that of energetic ion effect in ZFZF generation by energetic ion driven TAE. This is due to the non-cancelation of thermal plasma RS\&MX stress; that is,  in the eBAE case one has  $1-k^2_{\parallel}V^2_A\simeq 1$; while, as in the TAE cases, one has $1-k^2_{\parallel}V^2_A\simeq O(\epsilon)$. Thus, one may expect that the EE phase space nonlinearity, e.g., phase-locking induced convective transport of EEs and the self-consistent nonlinear evolution, could be a more important channel for eBAE nonlinear saturation \cite{FZoncaNJP2015,TWangPoP2019,MFalessiPoP2019}.

\section*{Acknowledgements}

This work is supported by   the National Key R\&D Program of China  under Grant Nos.  2017YFE0300501 and 2017YFE0301900,
the National Science Foundation of China under grant No.   11875233.
This work was also carried out within the framework of the EUROfusion
Consortium and received funding from the EURATOM research and training programme
2014 - 2018 and 2019 - 2020 under Grant Agreement No. 633053 (Project Nos. WP19-ER/ENEA-05 and
WP17-ER/MPG-01). The views and opinions expressed herein do not necessarily reflect
those of the European Commission. LC also acknowledges support of US DoE grant.

\end{document}